\documentclass[12pt,a4paper]{amsart}

\usepackage{amsmath,amsfonts,amssymb}

\usepackage[hmargin=2cm,vmargin=2cm]{geometry}

\usepackage{hyperref}
\usepackage{nameref,zref-xr}                    

\newcommand{\mbC}{\mathbb C}

\def\CP1{\mathbb{C}\mathrm{P}^1}

\newcommand{\oM}{\overline{\mathcal M}}

\newcommand{\oh}{{\overline h}}

\newcommand{\eps}{\varepsilon}
\def\d{\partial}

\newcommand{\<}{\left <}
\renewcommand{\>}{\right >}

\newcommand{\Coef}{\mathop{\mathrm{Coef}}\nolimits}

\newcommand{\tF}{\widetilde F}

\newcommand{\Hodge}{\mathrm{Hodge}}
\newcommand{\ILW}{\mathrm{ILW}}
\newcommand{\tS}{\widetilde S}

\newtheorem{theorem}{Theorem}[section]

\theoremstyle{definition}

\newtheorem{remark}[theorem]{Remark}

\numberwithin{equation}{section}

\title[ILW equation for Hodge integrals]{ILW equation for the Hodge integrals revisited}

\author{Alexandr Buryak}
\address{Department of Mathematics \\ ETH Zurich \\ R\"amistrasse 101 8092, Zurich, Switzerland}
\email{buryaksh@gmail.com}

\begin{document}

\begin{abstract}
In a previous paper we proved that after a simple transformation the generating series of the linear Hodge integrals on the moduli space of stable curves satisfies the hierarchy of the Intermediate Long Wave equation. In this paper we present a much shorter proof of this fact. Our new proof is based on an explicit formula for the one-point linear Hodge integrals that was found independently by Faber, Pandharipande and Ekedahl, Lando, Shapiro, Vainshtein.
\end{abstract}

\maketitle

\section{Introduction}

Consider the moduli space~$\oM_{g,n}$ of stable complex algebraic curves of genus~$g$ with~$n$ marked points. The class $\psi_i\in H^2(\oM_{g,n};\mbC)$ is defined as the first Chern class of the line bundle over~$\oM_{g,n}$ formed by the cotangent lines at the $i$-th marked point. Denote by $\lambda_j\in H^{2j}(\oM_{g,n};\mbC)$ the $j$-th Chern class of the rank~$g$ Hodge vector bundle over~$\oM_{g,n}$ whose fibers over smooth curves are the spaces of holomorphic one-forms. In~\cite{Bur15} we studied the following integrals:
\begin{gather}\label{eq:bracket}
\<\lambda_j\tau_{d_1}\tau_{d_2}\ldots\tau_{d_n}\>_g:=\int_{\oM_{g,n}}\lambda_j\psi_1^{d_1}\psi_2^{d_2}\ldots\psi_n^{d_n}.
\end{gather}
In the unstable cases $2g-2+n\le 0$ we define the bracket to be equal to zero. Introduce variables $\hbar,\eps$ and $t_0,t_1,t_2,\ldots$ and consider the generating series
$$
F^{\Hodge}(t_0,t_1,\ldots;\hbar,\eps):=\sum_{n\ge 0}\sum_{0\le j\le g}\frac{\hbar^g\eps^j}{n!}\sum_{d_1,\ldots,d_n\ge 0}\<\lambda_j\tau_{d_1}\ldots\tau_{d_n}\>_g t_{d_1}\ldots t_{d_n}.
$$
Let
\begin{gather*}
\tF^{\Hodge}:=F^{\Hodge}+\sum_{g\ge 1}\frac{(-1)^g}{2^{2g}(2g+1)!}\hbar^{g}\eps^g\frac{\d^{2g}F^{\Hodge}}{\d t_0^{2g}}.
\end{gather*}
In~\cite{Bur15} we proved the following theorem.
\begin{theorem}\label{theorem:main theorem}
The second derivative $u=\frac{\d^2 \tF^{\Hodge}}{\d t_0^2}$ satisfies the hierarchy of the Intermediate Long Wave equation:
\begin{align}
\frac{\d u}{\d t_1}&=u u_x+\sum_{g\ge 1}\hbar^g\eps^{g-1}\frac{|B_{2g}|}{(2g)!}u_{2g+1},\label{eq:ILW hierarchy}\\
\frac{\d u}{\d t_2}&=\frac{1}{2}u^2u_x+\sum_{g\ge 1}\frac{|B_{2g}|}{(2g)!}\hbar^g\frac{\eps^{g-1}}{4}(2(u u_{2g})_x+\d_x^{2g+1}(u^2))+\sum_{g\ge 2}\frac{|B_{2g}|}{(2g)!}\hbar^g\eps^{g-2}(g+1)u_{2g+1},\notag\\
&\vdots\notag
\end{align}
\end{theorem}
\noindent Here we identify $x$ with $t_0$, $B_{2g}$ stand for the Bernoulli numbers: $B_2=\frac{1}{6},B_4=-\frac{1}{30},\ldots$; and we denote by~$u_i$ the derivative~$\d_x^i u$. 

Let us say a few words about the hierarchy of the Intermediate Long Wave (ILW) equation (see e.g.~\cite{SAK79}), since it is less known in the mathematical literature than, for example, the KdV hierarchy. The ILW equation is actually the first equation~\eqref{eq:ILW hierarchy} of the hierarchy. It can be written in a hamiltonian form:
\begin{align*}
&\frac{\d u}{\d t_1}=\d_x\frac{\delta\oh^{\ILW}_1}{\delta u},\quad\text{where}\\
&\oh^{\ILW}_1=\int\left(\frac{u^3}{6}+\sum_{g\ge 1}\hbar^g\eps^{g-1}\frac{|B_{2g}|}{2(2g)!}u u_{2g}\right)dx.
\end{align*}
The whole ILW hierarchy is also hamiltonian. The higher Hamiltonians $\oh^{\ILW}_i$, $i\ge 2$, have the form
\begin{gather*}
\oh^{\ILW}_i=\int\left(\frac{u^{i+2}}{(i+2)!}+O(\hbar)\right)dx,
\end{gather*} 
and, by \cite[Lemma 2.4]{Bur15}, are uniquely determined by the commutation relation
$$
\left\{\oh^{\ILW}_i,\oh^{\ILW}_1\right\}_{\d_x}:=\int\frac{\delta\oh^{\ILW}_i}{\delta u}\d_x\frac{\delta\oh^{\ILW}_1}{\delta u}dx=0.
$$
An explicit construction of the higher Hamiltonians can be found in~\cite{SAK79}.

\begin{remark}
We must say that our definition of the ILW hierarchy is slightly different from the one that is more common in the physics literature (see e.g.~\cite{SAK79}). First of all, the ILW equation from~\cite{SAK79} is different from ours by several rescalings and, second, the higher Hamiltonians from~\cite{SAK79} are related to ours by some triangular transformation. We discuss the precise relation in~\cite{Bur15}.
\end{remark}

\begin{remark}
In~\cite{Bur15} the hierarchy~\eqref{eq:ILW hierarchy} is called the deformed KdV hierarchy. This happened because at the moment when the result of~\cite{Bur15} was obtained we didn't know that this hierarchy already appeared in a literature. We are very grateful to S.~Ferapontov and D.~Novikov for recognizing the ILW equation in the first equation of~\eqref{eq:ILW hierarchy} after the author's talk on the conference in Trieste (Hamiltonian PDEs, Frobenius manifolds and Deligne-Mumford moduli spaces, September 2013).
\end{remark}

Our approach in~\cite{Bur15} is based on the Dubrovin and Zhang theory generalized in~\cite{BPS12a} (see also~\cite{BPS12b}). Let us briefly recall it. Suppose we have a semisimple cohomological field theory (not necessarily homogeneous) with a metric $\eta$ and a potential~$F$. Then in~\cite{BPS12a} it is presented a construction of a hamiltonian hierarchy of PDEs with a solution given by the second derivatives~
$$
w^\alpha=\eta^{\alpha\gamma}\frac{\d^2 F}{\d t^1_0\d t^\gamma_0}.
$$
Consider the cohomological field theory formed by the full Chern class of the Hodge bundle:
\begin{gather}\label{Hodge classes}
1+\eps\lambda_1+\eps\lambda_2+\ldots+\eps^g\lambda_g\in H^*(\oM_{g,n};\mbC).
\end{gather}  
In~\cite{Bur15} we showed that Theorem~\ref{theorem:main theorem} is a corollary of the following result.
\begin{theorem}\label{theorem: DZ hierarchy}
Consider the Dubrovin-Zhang hierarchy associated to the cohomological field theory \eqref{Hodge classes}. Then the Miura transformation 
\begin{gather}\label{eq:miura}
w\mapsto u=w+\sum_{g\ge 1}\frac{(-1)^g}{2^{2g}(2g+1)!}\hbar^{g}\eps^g w_{2g}
\end{gather}
transforms this hierarchy to the ILW hierarchy. 
\end{theorem}

The proof of Theorem~\ref{theorem: DZ hierarchy} in~\cite{Bur15} consists of two steps:
\begin{enumerate}

\item Note that the class~\eqref{Hodge classes} lies in cohomological degrees not greater than~$2g$. Remarkably, this simple fact together with the polynomiality property of the Dubrovin-Zhang hierarchy, proved in~\cite{BPS12a} (see also \cite{BPS12b}), and an explicit formula for the $\lambda_g$-integrals, proved in~\cite{FP03}, imply that the Miura transformation~\eqref{eq:miura} transforms the first hamiltonian operator of the Dubrovin-Zhang hierarchy to the operator~$\d_x$ and the first Hamiltonian to a local functional of the form
\begin{gather}\label{form with unknown coefficients}
\int\left(\frac{u^3}{6}+\sum_{g\ge 1}\hbar^g\eps^{g-1}C_g u u_{2g}\right)dx,
\end{gather}
where $C_g$ are some constants.

\item The second step consists of the computation of the constants~$C_g$. In~\cite{Bur15} we observed that an existence of a local functional of the form
$$
\int\left(\frac{u^4}{24}+O(\hbar)\right)dx,
$$
that commutes with~\eqref{form with unknown coefficients}, gives strong constraints for the coefficients~$C_g$. This idea together with the explicit computation of the coefficients $C_1$ and~$C_2$ allowed us to prove that $C_g=\frac{|B_{2g}|}{2(2g)!}$. 

\end{enumerate}
While the first step is a direct computation (see \cite[Section 5]{Bur15}) based on the construction of the Dubrovin-Zhang hierarchy from~\cite{BPS12a}, the second step in~\cite{Bur15} involves quite technical computations (see \cite[Sections 6,7 and Appendix A and B]{Bur15}). 

In the present paper we do the second step in a much simpler way. We give a short computation of the constants~$C_g$ using the explicit formula for the one-point linear Hodge integrals that was found independently in~\cite{FP00} and~\cite{ELSV99}.

For a reader who is not familiar with details of the Dubrovin-Zhang theory we would like to make the following summary. In~\cite[Section 5]{Bur15} we proved that the second derivative $u=\frac{\d^2\tF^{\Hodge}}{\d t_0^2}$ satisfies an equation of the form
$$
\frac{\d u}{\d t_1}=uu_x+\sum_{g\ge 1}\hbar^g\eps^{g-1}2C_g u_{2g+1},
$$
for some constants $C_g$. In Section~\ref{section:computation} we present a short computation of the constants~$C_g$. This gives another proof of Theorem~\ref{theorem: DZ hierarchy}. The computation in Section~\ref{section:computation} doesn't use any material from~\cite{Bur15}. So it is possible to read Section~\ref{section:computation} even without looking inside~\cite{Bur15}. 

\begin{remark}
Our argument from Section~\ref{section:computation} can be easily applied in the opposite direction. It shows that the formula from~\cite{FP00} and~\cite{ELSV99} for the generating series of the one-point linear Hodge integrals (formula~\eqref{eq:one-point Hodge}) follows from Theorem~\ref{theorem:main theorem}.
\end{remark}

\subsection{Acknowledgements} 

We would like to thank R.~Pandharipande and P.~Rossi for useful discussions. 

The author was supported by grant ERC-2012-AdG-320368-MCSK in the group of R.~Pandharipande at ETH Zurich, by grants RFFI 13-01-00755 and NSh-4850.2012.1.
  
%%%%%%%%%%%%%%%%%%%%%%%%%%%%%%%%%%%%%%%%%%%%%%%%%%%%%%%%%%%%%%%%%
%%%%%%%%%%%%%%%%%%%%%%%%%%%%%%%%%%%%%%%%%%%%%%%%%%%%%%%%%%%%%%%%%

\section{Computation of the constants $C_g$}\label{section:computation}

We have the equation
\begin{gather}\label{eq:equation fot u}
\frac{\d u}{\d t_1}=u u_x+\sum_{g\ge 1}\hbar^g\eps^{g-1}2C_g u_{2g+1},
\end{gather}
where $u=\frac{\d^2\tF^{\Hodge}}{\d t_0^2}$ and $C_g$ are some constants. Let us prove that $C_g=\frac{|B_{2g}|}{2(2g)!}$.

Consider formal variables $t$ and $k$. The following formula was obtained independently in~\cite{FP00} and~\cite{ELSV99}:
\begin{gather}\label{eq:one-point Hodge}
1+\sum_{g\ge 1}\sum_{i=0}^g t^{2g}k^i\<\lambda_{g-i}\tau_{2g-2+i}\>_g=\left(\frac{t/2}{\sin(t/2)}\right)^{k+1}.
\end{gather}
Let us slightly reformulate this result. We have the string equation
\begin{gather}\label{eq:string equation}
\frac{\d F^{\Hodge}}{\d t_0}=\sum_{n\ge 0}t_{n+1}\frac{\d F^{\Hodge}}{\d t_n}+\frac{t_0^2}{2}+\frac{\hbar\eps}{24}.
\end{gather}
Note also that the bracket~\eqref{eq:bracket} vanishes unless the dimension constraint
\begin{gather}\label{eq:dimension constraint}
3g-3+n=j+\sum_{i=1}^n d_i
\end{gather}
is satisfied. Let 
$$
S(\hbar,\eps):=\sum_{0\le j\le g}\<\lambda_j\tau_0^2\tau_{3g-j}\>_g\hbar^g\eps^j.
$$
By the string equation~\eqref{eq:string equation}, formula~\eqref{eq:one-point Hodge} is equivalent to the following equation:
\begin{gather}\label{eq:reformulation}
S(\hbar,\eps)=\left(\frac{\sqrt{\hbar\eps}/2}{\sin(\sqrt{\hbar\eps}/2)}\right)^{1+\frac{1}{\eps}}.
\end{gather}
Let
\begin{gather}\label{eq:new bracket}
\<\lambda_j\tau_{d_1}\tau_{d_2}\ldots\tau_{d_n}\>^{\sim}_g:=\Coef_{\hbar^g\eps^j}\left.\frac{\d^n\tF^{\Hodge}}{\d t_{d_1}\d t_{d_2}\ldots \d t_{d_n}}\right|_{t_*=0}.
\end{gather}
It is easy to see that, similarly to the bracket~\eqref{eq:bracket}, the new bracket~\eqref{eq:new bracket} vanishes unless the dimension constraint~\eqref{eq:dimension constraint} is satisfied. It is also clear that the new bracket is zero when $j>g$. Let
$$
\tS(\hbar,\eps):=\sum_{0\le j\le g}\<\lambda_j\tau_0^2\tau_{3g-j}\>^{\sim}_g\hbar^g\eps^j.
$$
Using the string equation~\eqref{eq:string equation} we get
\begin{align*}
\tS(\hbar,\eps)=&\left(1+\sum_{g\ge 1}\frac{(-1)^g}{2^{2g}(2g+1)!}\hbar^g\eps^g\right)S(\hbar,\eps)=\frac{\sin(\sqrt{\hbar\eps}/2)}{\sqrt{\hbar\eps}/2}S(\hbar,\eps)\stackrel{\text{by~\eqref{eq:reformulation}}}{=}\\
=&\left(\frac{\sqrt{\hbar\eps}/2}{\sin(\sqrt{\hbar\eps}/2)}\right)^{1/\eps}=\exp\left(\frac{1}{\eps}\log\left(\frac{\sqrt{\hbar\eps}/2}{\sin(\sqrt{\hbar\eps}/2)}\right)\right).
\end{align*}
Note that
$$
\log\left(\frac{z/2}{\sin(z/2)}\right)=\sum_{g\ge 1}\frac{|B_{2g}|}{2g(2g)!}z^{2g}.
$$
Therefore,
\begin{gather}\label{eq:corollary of one-point}
\tS(\hbar,\eps)=\exp\left(\sum_{g\ge 1}\hbar^g\eps^{g-1}\frac{|B_{2g}|}{2g(2g)!}\right).
\end{gather}

Let us now compute~$\tS(\hbar,\eps)$ using equation~\eqref{eq:equation fot u}. The string equation~\eqref{eq:string equation} for~$F^{\Hodge}$ implies the string equation for $\tF^{\Hodge}$:
$$
\frac{\d\tF^{\Hodge}}{\d t_0}=\sum_{n\ge 0}t_{n+1}\frac{\d \tF^{\Hodge}}{\d t_n}+\frac{t_0^2}{2}.
$$
It gives the property $u_d|_{t_*=0}=\delta_{d,1}$. Therefore, we can compute the linear term in~$t_i$'s of the expression on the right-hand side of equation~\eqref{eq:equation fot u} in the following way:
\begin{align}
&u u_x+\sum_{g\ge 1}\hbar^g\eps^{g-1}2C_g u_{2g+1}=\label{eq:linear on the right}\\
=&\sum_{0\le j\le g}\hbar^g\eps^j\<\lambda_j\tau_0^2\tau_{3g-j}\>^\sim_g t_{3g-j}+\sum_{g\ge 1}2C_g\sum_{0\le j\le h}\hbar^{g+h}\eps^{g+j-1}\<\lambda_j\tau_0^2\tau_{3h-j}\>^\sim_h t_{3(g+h)-(g+j-1)}+O(t^2).\notag
\end{align}
Let us look on the left-hand side of equation~\eqref{eq:equation fot u}. We have the dilaton equation for the potential~$F^{\Hodge}$:
$$
\frac{\d F^{\Hodge}}{\d t_1}=\left(\sum_{n\ge 0}t_n\frac{\d}{\d t_n}+2\hbar\frac{\d}{\d\hbar}-2\right)F^{\Hodge}+\frac{\hbar}{24}.
$$
It implies the dilaton equation for the transformed potential~$\tF^{\Hodge}$:
$$
\frac{\d\tF^{\Hodge}}{\d t_1}=\left(\sum_{n\ge 0}t_n\frac{\d}{\d t_n}+2\hbar\frac{\d}{\d\hbar}-2\right)\tF^{\Hodge}+\frac{\hbar}{24}.
$$
This allows us to compute the linear term in~$t_i$'s of the expression on the left-hand side of~\eqref{eq:equation fot u} in the following way:
\begin{gather}\label{eq:linear on the left}
\frac{\d u}{\d t_1}=\sum_{0\le j\le g}\hbar^g\eps^j(2g+1)\<\lambda_j\tau_0^2\tau_{3g-j}\>_g^\sim t_{3g-j}+O(t^2).
\end{gather}
Equating~\eqref{eq:linear on the right} and~\eqref{eq:linear on the left}, we get
$$
\sum_{0\le j\le g}\hbar^g\eps^j(2g)\<\lambda_j\tau_0^2\tau_{3g-j}\>_g^\sim t_{3g-j}=\sum_{g\ge 1}2C_g\sum_{0\le j\le h}\hbar^{g+h}\eps^{g+j-1}\<\lambda_j\tau_0^2\tau_{3h-j}\>^\sim_h t_{3(g+h)-(g+j-1)}.
$$
We can rewrite this equation as a differential equation for~$\tS(\hbar,\eps)$:
\begin{gather}\label{eq:differerential equation}
\frac{\d}{\d\hbar}\tS(\hbar,\eps)=\left(\sum_{g\ge 1}C_g(\hbar\eps)^{g-1}\right)\tS(\hbar,\eps).
\end{gather}
Therefore,
$$
\tS(\hbar,\eps)=\exp\left(\sum_{g\ge 1}\frac{C_g}{g}\hbar^g\eps^{g-1}\right).
$$
Comparing this equation with equation~\eqref{eq:corollary of one-point}, we get
$$
C_g=\frac{|B_{2g}|}{2(2g)!}.
$$ 
This completes our proof.

\end{document}